# Transverse-electric Cherenkov Radiation for TeV-Scale Particle Detection

Zhixiong Xie[1], Xiao Lin[2,3], Song Zhu[1], Chunyu Huang[1], Yu Luo[1,*], and Hao Hu[1,*]

[1]National Key Laboratory of Microwave Photonics, College of Electronic and Information Engineering, Nanjing University of Aeronautics and Astronautics, Nanjing 211106, China

[2]Interdisciplinary Center for Quantum Information, State Key Laboratory of Extreme Photonics and Instrumentation, Zhejiang University, Hangzhou 310027, China

[3]International Joint Innovation Center, The Electromagnetics Academy at Zhejiang University, Zhejiang University, Haining 314400, China

*Corresponding authors: Yu Luo, Hao Hu
**Email:** yu.luo@nuaa.edu.cn (Yu Luo); hao.hu@nuaa.edu.cn (Hao Hu)

**Cherenkov radiation enables high-energy particle identification through its velocity-dependent emission angle, yet conventional detectors fail to detect momenta beyond tens of GeV/$c$ owing to the absence of natural materials with near-unity refractive indices. We overcome this limitation by demonstrating directional Cherenkov radiation from transverse-electric (TE) graphene plasmons, excited by a swift charged particle travelling above suspended monolayer graphene. Crucially, TE graphene plasmons exhibit a near-unity mode index, sustaining high sensitivity of the Cherenkov angle to relativistic velocities up to the TeV/$c$ regime. The radiation further maintains exceptional robustness against particle-graphene separation changes, enabled by the TE mode's low transverse decay rate. This ultracompact platform is electrically tunable, allowing on-chip, reconfigurable detection of ultrahigh-energy particles and extending measurable momenta by two orders of magnitude beyond existing detectors.**

## 1. Introduction

Cherenkov radiation, first experimentally discovered by P. A. Cherenkov in 1934,[1] is an electromagnetic radiation phenomenon wherein photons are emitted by a charged particle traveling faster than the phase velocity of light in a transparent medium. Cherenkov radiation is highly directional, forming a characteristic conical emission pattern aligned with the particle trajectory. According to Frank and Tamm's theory,[2,3] the emission angle $\theta$ (known as the Cherenkov angle) satisfies $\cos\theta = c/nv_e$ in homogeneous isotropic media, where $c$ is the light speed in vacuum, $v_e$ is the particle velocity and $n$ is the refractive index of the medium. The dependence of this angle on the particle velocity forms the basis of Cherenkov detectors, leading to the discovery of many elementary particles such as anti-protons and $J/\psi$ particles.[4–6]

However, the performance of conventional Cherenkov detectors is fundamentally constrained by the refractive index of the host materials. Accurate identification of high-momentum particles (i.e., with $v_e \to c$) requires host materials with a refractive index decreasingly close to unity (i.e., $n \to 1$).[7–11] For instance, silicon aerogel detectors, whose host material with different structures typically has a refractive index ranging from 1.005 to 1.060, can effectively discriminate particles only up to momenta around 10 GeV/$c$.[8,11–13] Further increase of particle momentum makes the Cherenkov angle insensitive to particle velocity, rendering traditional detectors ineffective. Therefore, overcoming this intrinsic material constraint is essential for enhancing the capabilities of Cherenkov-based particle detection.

A variety of modern material technologies such as metamaterials and photonic crystals have been proposed to relax the aforementioned material limitations and to enhance the sensitivity of Cherenkov detectors. For instance, using the framework of transformation optics, anisotropic metamaterials have been designed to control the Cherenkov angle for particles with momenta in the range of 30–100

GeV/$c$.[14] To mitigate the loss typically associated with conventional metamaterials, a solid composite invisible material has been developed that achieves ideal electromagnetic transparency by regulating its internal polarization properties.[15,16] Meanwhile, resonance transition radiation in photonic crystals has been introduced, demonstrating a strong sensitivity to particle momentum above 500 GeV/$c$.[17] To further broaden the bandwidth, a Cherenkov detector with lossless, dispersionless, and highly sensitive characteristics has been proposed utilizing the Brewster effect in cascaded photonic crystals.[18] Despite these promising developments, most approaches rely on complex multilayered composite structures, requiring precise customization of geometric and material parameters for each constituent layer. These intricate designs significantly complicate experimental implementation, posing substantial challenges to the practical realization of highly sensitive Cherenkov detectors.

Recent studies have demonstrated that plasmonic/phononic structures offer an experimentally feasible platform for controlling Cherenkov radiation.[19–25] For example, two-dimensional (2D) Cherenkov radiation has been experimentally observed on a metallodielectric plasmonic structure, showing quantum coupling strength over 2 orders of magnitude larger than previous work.[26] More recent work reports reversed 2D Cherenkov radiation mediated by hyperbolic phonon polaritons with negative group velocity.[27] While 2D Cherenkov radiation holds promise for the development of integrated free-electron light sources, its potential application in Cherenkov detectors has received limited attention. The primary challenge lies in the nature of most conventional transverse-magnetic (TM) surface plasmon or photon polaritons, which are highly confined and exhibit mode refractive indices significantly greater than unity.[28–36] As a result, rather than enhancing sensitivity, 2D Cherenkov radiation has been widely regarded as detrimental to detector performance. To date, whether 2D Cherenkov radiation can be effectively utilized for high-energy particle detection remains an open question.

To this end, we propose a novel type of 2D Cherenkov radiation mediated by low-index plasmons, as exemplified by TE graphene plasmons, enabling sensitivity enhancement of Cherenkov angle to the relativistic particle velocity. TE graphene plasmons are known as collective oscillations of charges localized near graphene layers, where the electric field vectors are in-plane (in a plane formed by the wavevector direction and the interface normal) and the magnetic field vector lies out of plane.[37–45] This is different from TM graphene plasmons where the magnetic field vectors are in-plane, while the electric field vector is out-of-plane. First introduced by S. A. Mikhailov and K. Ziegler in 2007,[46] TE graphene plasmons can propagate along graphene layers within frequency bands dominated by interband optical transitions. Notably, their mode refractive indices are extremely close to unity, making TE graphene plasmons a potential platform for constructing surface Cherenkov detectors. Here, we study the emission behaviors of a swift charged particle moving atop a suspended graphene monolayer. Our calculations reveal that the emission angle of TE graphene plasmons remains highly sensitive to particle velocity, even for momenta exceeding 10 GeV/$c$. Furthermore, by tuning the chemical potential of graphene, we demonstrate that the detectable momentum range can be extended to several TeV/$c$, over two orders of magnitude beyond the limits of conventional bulk Cherenkov detectors[11–13] based on silica aerogel or gas. Another key advantage of TE graphene plasmon excitation lies in its robustness with respect to the particle-graphene separation. Specifically, the excitation efficiency remains at a high level even at separation distances exceeding 100 nm, showing over an order-of-magnitude efficiency enhancement compared to TM graphene plasmon excitation. This robustness stems from the negligible transverse decay rate of TE graphene plasmons. In sharp contrast, the efficient excitation of TM graphene plasmons requires the distance between the particle trajectory and the graphene sheet down to a few nanometers, posing significant experimental challenges.

## 2. Results and discussion

Without loss of generality, we consider a swift charged particle moving in vacuum (with the relative permittivity of $\varepsilon_1$) and atop a suspended monolayer graphene (see the schematic in Figure 1a). The current density of the charged particle is $\vec{J}(\vec{r},t) = \hat{z}v_e e\delta(x)\delta(y-y_0)\delta(z-v_e t)$, where $\vec{v}_e = \hat{z}v_e$ is the particle velocity and $e$ is the elementary charge. The suspended monolayer graphene is located at $y = 0$, with a surface conductivity of $\sigma_g$ and a separation distance from the particle trajectory of $y_0$. Kubo formula is applied to describe the graphene conductivity, with chemical potential and relaxation time denoted as $\mu_c$ and $\tau$, respectively. To radiate excited graphene plasmons, we adopt the silicon dioxide (SiO$_2$) with the relative permittivity of $\varepsilon_2$ as the substrate. The thickness of suspended layer between the graphene sheet and SiO$_2$ substrate is $d$. The gold electrodes in contact with suspended graphene are used to modulate gate voltage of the graphene. Without particular statement, the adopt parameters are as followings: $\varepsilon_1 = 1$, $\varepsilon_2 = 2.4$, $y_0 = 5$ nm, $\mu_c = 0.145$ eV, $\tau = 0.1$ ps and $d = 200$ μm.

Our structure supports Cherenkov radiation mediated by TE graphene plasmons. To demonstrate this, we calculate the dispersion of TE graphene plasmons in our structure. By enforcing the boundary conditions for TE waves, the dispersion relation of TE graphene plasmons is obtained as

$$\left(\frac{2k_{y_1}+\sigma_g\omega\mu_0}{-\sigma_g\omega\mu_0}\right)\left(\frac{k_{y_1}+k_{y_2}}{k_{y_1}-k_{y_2}}\right) = e^{2ik_{y_1}d} \tag{1}$$

where $k_{y_j} = \sqrt{\varepsilon_j k_0^2 - q^2}$ is the $y$ component of wavevector in vacuum ($j = 1$) and substrate region ($j = 2$), $k_0 = \omega/c$ is the wavevector in free space, $q = \sqrt{k_x^2 + k_z^2}$ is the in-plane wavevector and $k_x$ ($k_z$) is the $x$ ($z$) component of wavevector, $\omega = 2\pi f$ is the angular frequency, and $\mu_0$ is the permeability in free space. Cherenkov radiation mediated by TE graphene plasmons arises when the phase-matching conditions are completed, i.e., $\bar{q}\bar{v}_e = \omega$. That is, the radiation excitation occurs at two point degeneracies (red circles) between the particle surface (red surface) and the isofrequency contour (black dashed line) in momentum space in Figure 1b.

Owing to the close-to-unity refractive index of TE graphene plasmons, Cherenkov radiation demonstrated here possesses a near-light-speed velocity threshold. Figure 1c compares the dispersion relations of TE and TM graphene plasmons. The dispersions reflect that TE graphene plasmons exist in the frequency band $1.667 < \hbar\omega/\mu_c < 2$ where $\mathrm{Im}(\sigma_g) < 0$, whereas TM graphene plasmons exist in the frequency band $0 < \hbar\omega/\mu_c < 1.667$ where $\mathrm{Im}(\sigma_g) > 0$. Here, $\hbar$ is the reduced Planck constant and $\mathrm{Im}(\sigma_g)$ is the imaginary part of the graphene conductivity $\sigma_g$. We note that the mode refractive index of TE graphene plasmons at $f_0 = 70$ THz is $n = q/k_0 = 1.0014$. This value is much smaller than that of TM graphene plasmons (oftentimes greater than 10) in their working frequency band. Since TE graphene plasmon investigated in this work possesses a small mode refractive index, the nonlocal effect can be reasonably neglected here.[28,47] Hence, the velocity threshold of TE graphene plasmon Cherenkov radiation is $v_{th} = c/n = 0.9986c$ at $f_0$ (see the blue circle in Figure 1d). On the contrary, velocity thresholds of TM graphene plasmons are generally one-order-of-magnitude smaller than those of TE graphene plasmons.

The Fourier spectrum of TE graphene plasmon Cherenkov radiation is sensitive to relativistic particle velocity. To illustrate this point, Figure 2 plots the Fourier spectrum of Cherenkov radiation produced by a charged particle with velocity higher than the velocity threshold. If the normalized velocity of charged particle is $\beta = 0.9999$, $0.9994$, and $0.9989$, the emission angle is $\theta = 2.808°$, 2.144°, and 1.146°, respectively. Here, $\beta = v_e/c$ is the particle velocity normalized by the light speed in vacuum. The Fourier spectrum reflects that the maximum magnitude of radiation occurs only when the wavevector of charged particle is phase-matching with that of TE graphene plasmons, i.e., $k_0/\beta = k_z$. Then the swift charged particle emits TE graphene plasmons with $|k_y| = \sqrt{k_0^2 - q^2}$ into the angle $\theta = \arccos(k_z/q)$. As the particle velocity increases, the altered particle wavevector makes the propagation angle of TE graphene plasmons decrease.

In addition to the Fourier spectrum of Cherenkov radiation, the velocity-dependent emission of TE graphene plasmon Cherenkov radiation is also reflected in the angular power spectral density. The angular power spectral density as a function of $\beta$ and $\theta$ in Figure 3a exhibits a sharp angle-dependent energy enhancement due to the excitation of TE graphene plasmons. As the normalized particle velocity varies from 0.9999, 0.9994, to 0.9989, the emission angle that leads to the maximum sharp energy enhancement is sensitively changed from 2.808°, 2.144°, to 1.146°. To facilitate the practical implementation, we also consider the influence of material loss (e.g., the relaxation time $\tau$ of graphene) on the angular power spectral density. Our results show that decreasing the relaxation time (i.e., increasing the dissipation loss) broadens the angular line width. Even so, the angular power spectral density still shows a high angular resolution for the relativistic particle, even if the relaxation time is significantly reduced (e.g., down to 0.025 ps).

TE graphene plasmon Cherenkov radiation could be exploited to identify high-energy particles, showing higher sensitivity as compared to conventional approaches. For example, deriving from the angle-velocity relation of Figure 3a, we show the relation between the particle momentum and the Cherenkov angle for four types of particles (i.e., electron, pion, kaon, and proton) in the solid lines of Figure 4. The results show that Cherenkov angles fixed at a momentum of 25 GeV/$c$ are 2.8601°, 2.8452°, 2.7070°, and 2.0282° for an electron, a pion, a kaon and a proton, respectively. Such a variation in $\theta$ indicates that different elementary particles with a momentum less than 50 GeV/$c$ can be effectively distinguished from one another if the chemical potential of graphene is adopted as $\mu_c =$ 0.145 eV. To further expand the detection range, we alter the $\mu_c$ which affects the conductivity of the graphene surface, as shown by dashed lines in Figure 4. When $\mu_c = 0.155$ eV, our method could distinguish particles with high momenta close to 0.3 TeV/$c$ (see the short dashed lines). More excitingly, the usage of $\mu_c = 0.165$ eV gives rise to a higher detection ability with the momentum larger than 5 TeV/$c$ (see the long dashed lines). Such a working momentum is over two orders of magnitude beyond the upper limits for conventional Cherenkov detectors (see comparison of the present method with conventional detectors using aerogel materials in the inset). In other words, the performance of proposed detectors could be readily enhanced by modulating gate voltage on the suspended graphene, without the reconfiguration of materials and structures.

Finally, the proposed TE graphene plasmon Cherenkov radiation exhibits strong robustness against the variations of particle-graphene separation $y_0$, owing to the long penetration depth of TE graphene plasmons. To highlight this point, we compare the intensities of TE and conventional TM graphene plasmon Cherenkov radiation as a function of $y_0$ (Figure 5). Our calculations reflect that the maximum magnitude of TE graphene plasmon Cherenkov radiation remains almost unchanged when $y_0$ increases from 5 nm to 50 nm, while that of TM graphene plasmons drastically decays by 84% (from $8.2 \times 10^{-14}$ J to $1.3 \times 10^{-14}$ J). To be specific, in the nanoscale separation (e.g., $y_0 < 5$ nm), the TE graphene plasmon Cherenkov radiation shows a slightly smaller emission efficiency than that of TM graphene plasmons. However, the peak-intensity of the former one suddenly surpasses the latter one as the separation distance increases to $y_0 = 9$ nm. More strikingly, in the large separation (e.g., $y_0 >$ 50 nm), TE graphene plasmon Cherenkov radiation exhibits strong superiority in the excitation efficiency, with the peak-intensity one-order-of-magnitude higher than that of TM graphene plasmon Cherenkov radiation. Without resorting to the nanoscale coupling distance, the potential collision between swift charged particles and samples could be favorably avoided in the practical excitation of TE graphene plasmon Cherenkov radiation[48,49].

## 3. Conclusion

In summary, our work enriches the family of 2D Cherenkov radiation by revealing TE graphene plasmon Cherenkov radiation in a suspended graphene structure. Owing to the close-to-unity mode index of TE graphene plasmons, the emission direction of Cherenkov radiation is susceptible to the particle velocity in the relativistic regime. The demonstrated radiation characteristics are suitable for high-energy particle identification in wide momentum ranges, with the upper limit over 5 TeV/$c$ which is unattainable by conventional particle detectors. Furthermore, the TE graphene plasmon Cherenkov radiation exhibits remarkable resilience to variations in the particle-graphene separation, significantly easing the stringent requirements typically associated with near-field charged particle excitation in graphene plasmons. We would like to emphasize that our proposed platform is general, as it can be realized not only via TE graphene plasmons but also through TM surface plasmons in ultrathin metallic slabs or TM surface phonon polaritons in thin-film polar dielectrics, thereby enabling operation across a broad frequency range (see more discussions in Section S6, Supplementary Information). On the other hand, although the high velocity threshold of the presented TE graphene plasmon Cherenkov radiation prevents its application from integrated free-electron light sources, the velocity threshold could be lowered down by adopting, e.g., negative-index metamaterials, as the substrate.[38,50] Our findings thus not only pave a feasible avenue to enable on-chip detection of relativistic particles, but also inspire future realization of novel free-electron light sources with controllable polarizations.

## Supporting Information
Supporting Information is available from the Wiley Online Library or from the author.

## Acknowledgements
This work is partially sponsored by National Natural Science Foundation of China (Grants No. 12404363); Natural Science Foundation of Jiangsu Province (Grants No. BK20241374); Distinguished Professor Fund of Jiangsu Province; Fundamental Research Funds for the Central Universities, NUAA (Grants No. NS2024022).

## Conflict of Interest
The authors declare no conflict of interest.

## Data Availability Statement
The data that support the findings of this study are available from the corresponding author upon reasonable request.

## References
[1]   P. A. Cherenkov, *Dokl. Akad. Nauk SSSR* **1934**, 2, 451.
[2]   I. M. Frank, I. Tamm, *Dokl. Akad. Nauk SSSR* **1937**, 14, 109.
[3]   I. M. Frank, *Science* **1960**, 131, 702.
[4]   O. Chamberlain, E. Segrè, C. Wiegand, T. Ypsilantis, *Phys. Rev.* **1955**, 100, 947.
[5]   J. J. Aubert, U. Becker, P. J. Biggs, J. Burger, M. Chen, G. Everhart, P. Goldhagen, J. Leong, T. McCorriston, T. G. Rhoades, M. Rohde, S. C. C. Ting, and S. L. Wu, Y. Y. Lee, *Phys. Rev. Lett.* **1974**, 33, 1404.
[6]   J.-E. Augustin, A. M. Boyarski, M. Breidenbach, et al. *Phys. Rev. Lett.* **1974**, 33, 1406.
[7]   T. Ypsilantis, J. Seguinot, *Nucl. Instrum. Meth. A* **1994**, 343, 30.
[8]   E. Nappi, *Nucl. Phys. B (Proc. Suppl.)* **1998**, 6, 270.
[9]   A. Abashian, K. Gotow, N. Morgan, et al. *Nucl. Instrum. Meth. A* **2002**, 479, 117.
[10]  A. A. Alves Jr, L. M. Andrade Filho, A. F. Barbosa, et al. (LHCb Collaboration) *J. Instrum.* **2008**, 3, S08005.
[11]  M. Adinolf, G. Aglieri Rinella, E. Albrecht, et al. (LHCb RICH Collaboration) *Eur. Phys. J. C* **2013**, 73, 2431.
[12]  S. S. Kistler, *Nature* **1931**, 127, 741.
[13]  A. C. Pierre, G. M. Pajonk, *Chem. Rev.* **2002**, 102, 4243.
[14]  V. Ginis, J. Danckaert, I. Veretennicoff, P. Tassin, *Phys. Rev. Lett.* **2014**, 113, 167402.
[15]  D. Ye, L. Lu, J. D. Joannopoulos, M. Soljačić, L. Ran, *Proc. Natl. Acad. Sci. U. S. A.* **2016**, 113, 2568.
[16]  C. Wang, X. Hu, L. Peng, J. Tang, L. Ran, S. Zhang, D. Ye, *Adv. Mater.* **2024**, 36, 2308298.
[17]  X. Lin, S. Easo, Y. Shen, H. Chen, B. Zhang, J. D. Joannopoulos, M. Soljačić, I. Kaminer, *Nat. Phys.* **2018**, 14, 816.
[18]  X. Lin, H. Hu, S. Easo, Y. Yang, Y. Shen, K. Yin, M. P. Blago, I. Kaminer, B. Zhang, H. Chen, J. Joannopoulos, M. Soljačić, Y. Luo, *Nat. Commun.* **2021**, 12, 5554.

[19] S. Xi, H. Chen, T. Jiang, L. Ran, J. Huangfu, B.-I. Wu, J. A. Kong, M. Chen, *Phys. Rev. Lett.* **2009**, 103, 194801.

[20] H. Chen, M. Chen, *Mater. Today* **2011**, 14, 34.

[21] P. Genevet, D. Wintz, A. Ambrosio, A. She, R. Blanchard, F. Capasso, *Nat. Nanotechnol.* **2015**, 10, 804.

[22] Z. Duan, X. Tang, Z. Wang, Y. Zhang, X. Chen, M. Chen, Y. Gong, *Nat. Commun.* **2017**, 8, 14901.

[23] H. Hu, X. Lin, Y. Luo, *Prog. Electromagn. Res.* **2021**, 171, 75.

[24] F. Liu, L. Xiao, Y. Ye, M. Wang, K. Cui, X. Feng, W. Zhang, Y. Huang, *Nat. Photonics* **2017**, 11, 289.

[25] H. Peng, T. W. Huang, K. Jiang, R. Li, C. N. Wu, M. Y. Yu, C. Riconda, S. Weber, C. T. Zhou, S. C. Ruan, *Phys. Rev. Lett.* **2023**, 131, 145003.

[26] Y. Adiv, H. Hu, S. Tsesses, R. Dahan, K. Wang, Y. Kurman, A. Gorlach, H. Chen, X. Lin, G. Bartal, I. Kaminer, *Phys. Rev. X* **2023**, 13, 011002.

[27] X. Guo, C. Wu, S. Zhang, D. Hu, S. Zhang, Q. Jiang, X. Dai, Y. Duan, X. Yang, Z. Sun, S. Zhang, H. Xu, Q. Dai, *Nat. Commun.* **2023**, 14, 2532.

[28] C. Wang, X. Chen, Z. Gong, R. Chen, H. Hao, H. Wang, Y. Yang, L. Tony, B. Zhang, H. Chen, X. Lin, *Rep. Prog. Phys.* **2024**, 87, 126401.

[29] S. Liu, P. Zhang, W. Liu, S. Gong, R. Zhong, Y. Zhang, M. Hu, *Phys. Rev. Lett.* **2012**, 109, 153902.

[30] A. N. Grigorenko, M. Polini, K. S. Novoslov, *Nat. Photonics* **2012**, 6, 749.

[31] G. X. Ni, A. S. Mcleod, Z. Sun, L. Wang, L. Xiong, K. W. Post, S. S. Sunku, B.-Y. Jiang, J. Hone, C. R. Dean, M. M. Fogler, D. N. Basov, *Nature* **2018**, 557, 530.

[32] Q. Xu, T. Ma, M. Danesh, B. N. Shivananju, S. Gan, J. Song, C. W. Qiu, H. M. Cheng, W. Ren, Q. Bao, *Light Sci. Appl.* **2017**, 6, e16204.

[33] F. Tay, X. Lin, X. Shi, H. Chen, I. Kaminer, B. Zhang, *Adv. Sci.* **2023**, 10, 2300760.

[34] R. Chen, Z. Gong, Z. Wang, X. Xi, B. Zhang, Y. Yang, B. Zhang, I. Kaminer, H. Chen, X. Lin, *Sci. Adv.* **2025**, 11, eads5113.

[35] Y. Luo, J. Zhao, A. Fieramosca, Q. Guo, H. Kang, X. Liu, T. C. H. Liew, D. Sanvitto, Z. An, S. Ghosh, Z. Wang, H. Xu, Q. Xiong, *Light Sci. Appl.* **2024**, 13, 203.

[36] Y. Shi, Y. Gan, Y. Chen, Y. Wang, S. Ghosh, A. Kavokin, Q. Xiong, *Nat. Mater.* **2025**, 24, 56.

[37] H. Hu, X. Lin, D. Liu, H. Chen, B. Zhang, Y. Luo, *Adv. Sci.* **2022**, 9, 2200538.

[38] V. G. Veselago, The electrodynamics of substances with simultaneously negative values of permittivity and permeability. *Sov. Phys. Usp.* **1968**, 10, 509.

[39] S. A. Maier, *Plasmonics: Fundamentals and Applications*, Springer, New York, NY **2007**.

[40] S. Chen, P. L. Leng, A. Konečná, E. Modin, M. Gutierrez-Amigo, E. Vicentini, B. Martín-García, M. Barra-Burillo, I. Niehues, C. Maciel Escudero, X. Y. Xie, L. E. Hueso, E. Artacho, J. Aizpurua, I. Errea, M. G. Vergniory, A. Chuvilin, F. X. Xiu, R. Hillenbrand, *Nat. Mater.* **2023**, 22, 860.

[41] Y. M. Qing, Y. Wang, Z. Yang, J. Wu, S. Yu, *Phys. Rev. A* **2024**, 109, 013504.

[42] Z. Ahmad, S. S. Oh, E. A. Muljarov, *Phys. Rev. Res.* **2024**, 6, 023185.

[43] Z. Xu, S. Bao, J. Liu, J. Chang, X. Kong, V. Galdi, T. J. Cui, *Laser Photonics Rev.* **2024**, 18, 2300763.

[44] Q. Ye, J Wang, Z. Liu, Z.-C. Deng, X.-T. Kong, F. Xing, X.-D. Chen, W.-Y. Zhou, C.-P. Zhang, J.-G. Tian, *Appl. Phys. Lett.* **2013**, 102, 021912.

[45] Y. V. Bludov, D. A. Smirnova, Y. S. Kivshar, N. M. R. Peres, M. I. Vasilevskiy, *Phys. Rev. B* **2014**, 89, 035406.

[46] S. A. Mikhailov, K. Ziegler, *Phys. Rev. Lett.* **2007**, 99, 016803.

[47] G. W. Hanson, *J. Appl. Phys.* **2008**, 103, 064302.

[48] P. Broaddus, T. Egenolf, D. S. Black, M. Murillo, C. Woodahl, Y. Miao, U. Niedermayer, R. L. Byer, K. J. Leedle, O. Solgaard, *Phys. Rev. Lett.* **2024**, 132, 085001.

[49] Z. Y. Liu, Q. Q. Wang, C. Y. Yang, K. Chen, D. D. Li, B. H. Lu, P. X. Yang, Y. D. Yang, *Opt. Express* **2025**, 33, 14737.

[50] X. Zhang, H. Hu, X. Lin, L. Shen, B. Zhang, H. Chen, *npj 2D Mater. Appl.* **2020**, 4, 25.

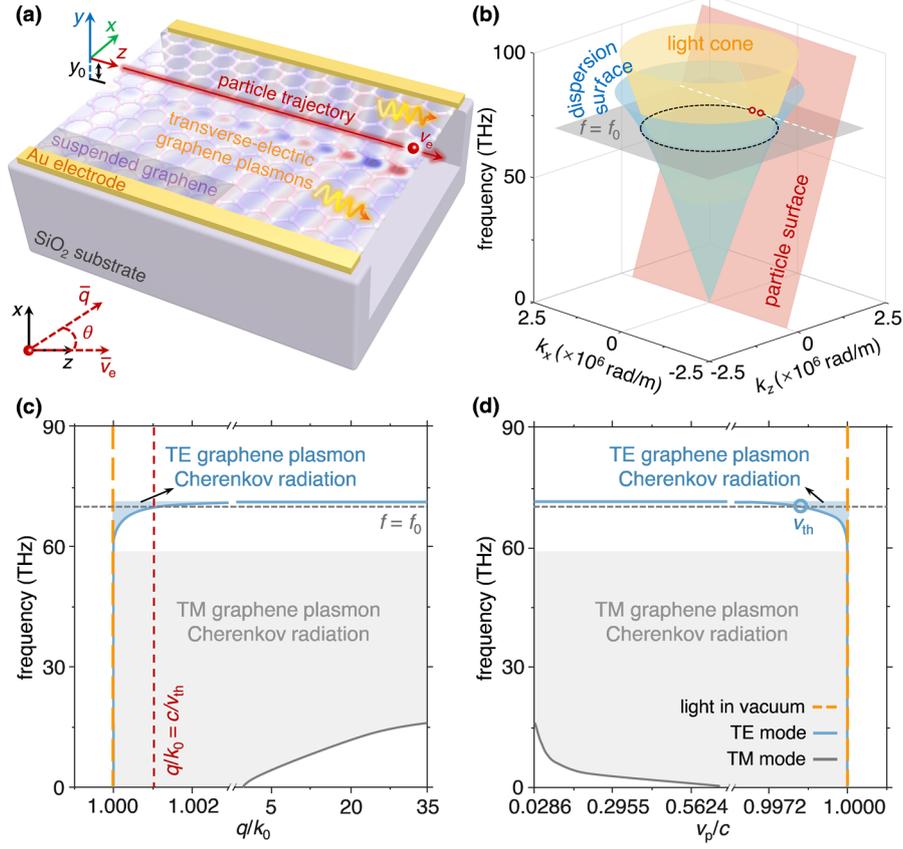

**Figure 1. Schematic of TE graphene plasmon Cherenkov radiation.** (a) Structural setup. A swift charged particle travels in vacuum parallel to the surface of a suspended graphene structure at a velocity $\vec{v}_e = \hat{z}v_e$, with a separation distance $y_0$ from the graphene sheet. The monolayer graphene is separated from the silicon dioxide ($SiO_2$) substrate by a suspended layer and is in contact with gold electrodes. The relative permittivities of the vacuum and $SiO_2$ are respectively denoted as $\varepsilon_1$ and $\varepsilon_2$. (b) Dispersion surfaces. The light cone, dispersion surface of TE graphene plasmons and particle surface are highlighted in yellow, blue and red surfaces, respectively. The intersections between the dispersion surface and particle surface correspond to the frequency and momentum of TE graphene plasmons excited by the swift charged particle. (c) Cutoff frequencies of TE and TM graphene plasmon Cherenkov radiation. (d) Velocity threshold of TE and TM graphene plasmon Cherenkov radiation. In (c, d), the existing domains for TE and TM graphene plasmon Cherenkov radiation are remarked in bule and gray shaded parts, respectively. Without particular statement, we adopt the following parameters here and below: the particle-graphene separation $y_0 = 5$ nm; the thickness of suspended layer $d = 200$ μm; the chemical potential of graphene $\mu_c = 0.145$ eV and the relaxation time of graphene $\tau = 0.1$ ps.

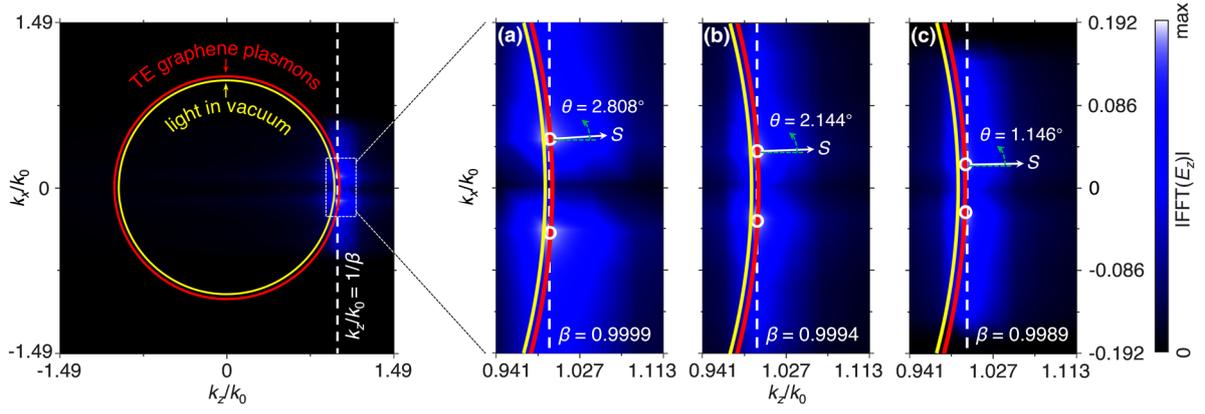

**Figure 2. Fourier spectrum of TE graphene plasmon Cherenkov radiation.** (a-c) The studied particle velocities are $\beta = 0.9999$, 0.9994 and 0.9989, respectively, leading to the emission angle of 2.808°, 2.144° and 1.146°, respectively. Here, the particle velocity is normalized by the light speed in vacuum, i.e., $\beta = v_e/c$, and the white arrows indicate the directions of Poynting vector $S$. The red and yellow solid curves represent the isofrequency contours of TE graphene plasmons and light in vacuum, respectively. The white dashed lines represent particle wavevectors, i.e., $k_z/k_0 = 1/\beta$. The white circles in (a)-(c) denote the wavevectors of excited TE graphene plasmon Cherenkov radiation.

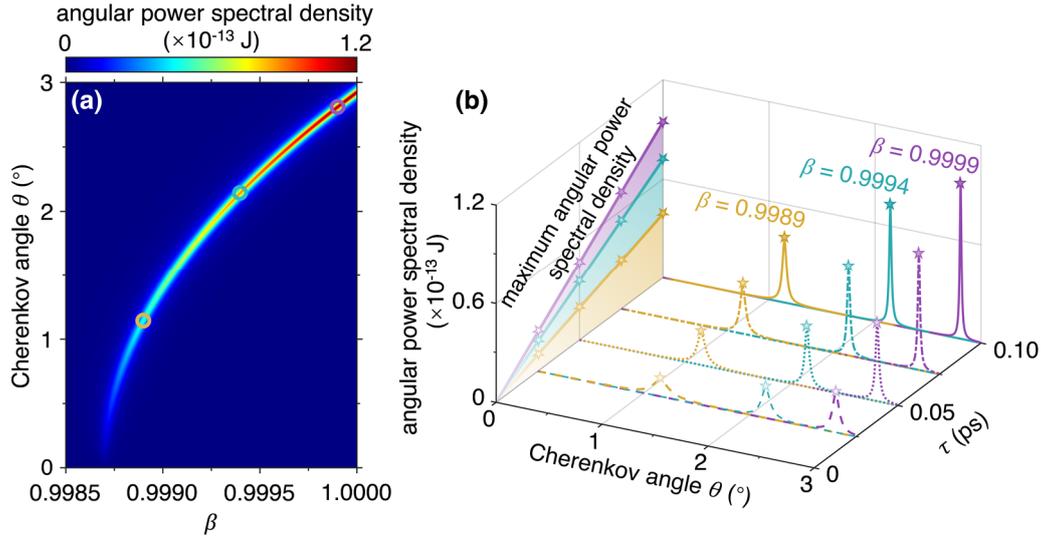

**Figure 3. Energy loss of a swift charged particle emitting TE graphene plasmons.** (a) The angular power spectral density as a function of the normalized particle velocity $\beta$ and the Cherenkov angle $\theta$ with $\tau = 0.1$ ps. (b) Influence of the relaxation time on the angular power spectral density. The studied particle velocities are $\beta = 0.9989$, 0.9994 and 0.9999 (as indicated by the yellow, green, and purple marks in (a), respectively). The studied relaxation times $\tau$ of graphene are 0.1 ps, 0.075 ps, 0.05 ps and 0.025 ps, respectively. The maximum angular power spectral densities as a function of relaxation time $\tau$ are highlighted in gradient yellow, green and purple lines.

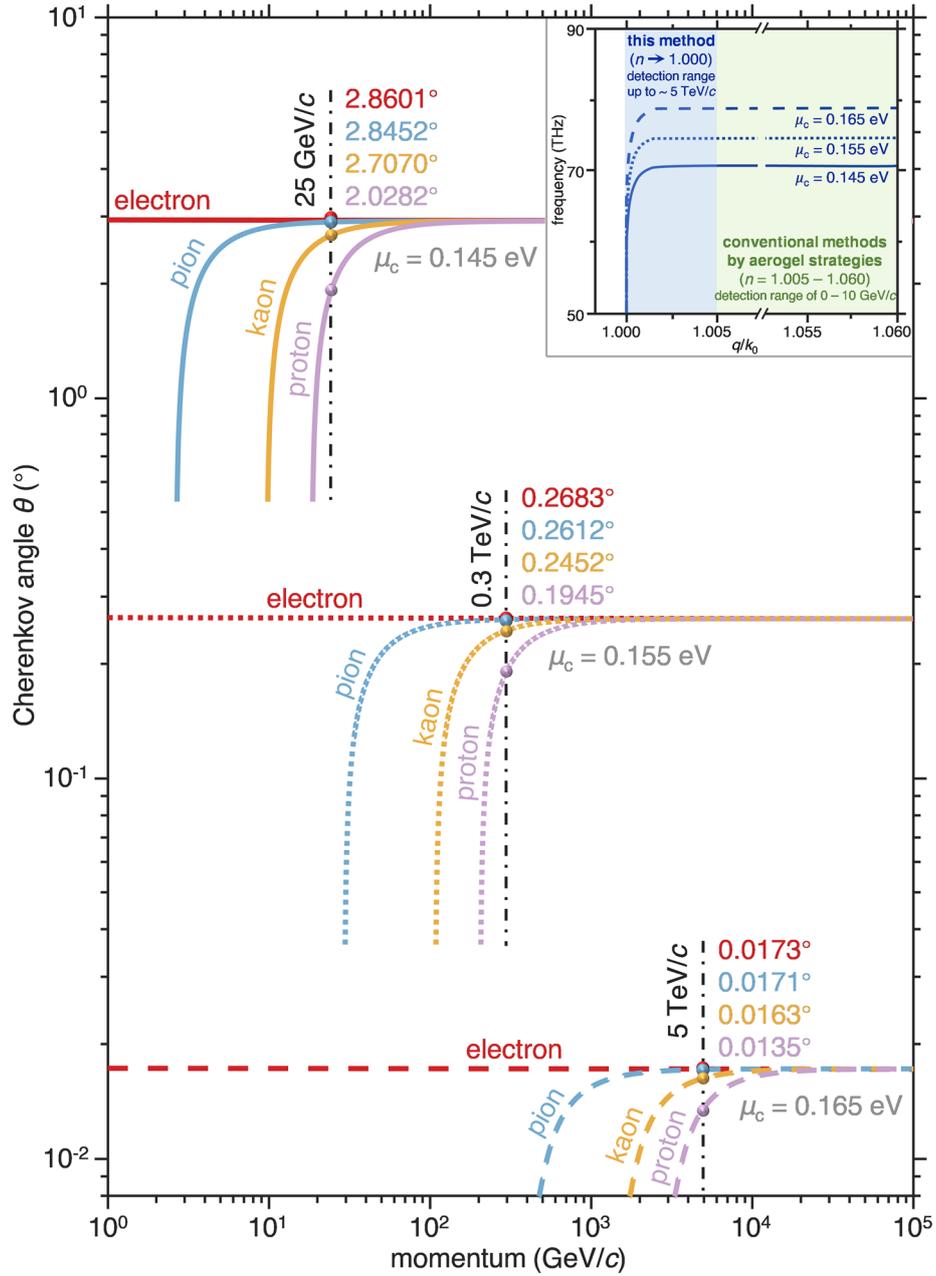

**Figure 4. Performance of particle detection with TE graphene plasmon Cherenkov radiation.** Cherenkov angles $\theta$ versus the particle momenta for four elementary particles: electron (red), pion (blue), kaon (yellow), and proton (purple). The inset plots the influence of the chemical potential on the dispersion curve of TE graphene plasmons, and comparison of performance between this method and conventional method using aerogel.

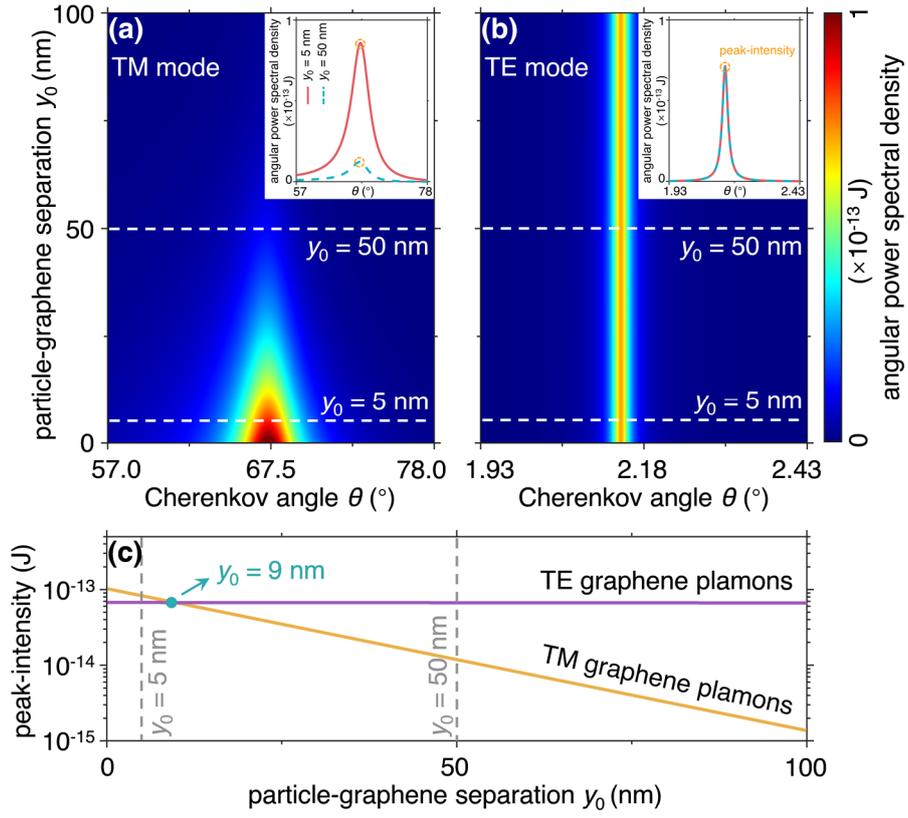

**Figure 5. Influence of the particle-graphene separation $y_0$ on the emission intensity of graphene plasmon Cherenkov radiation.** (a) The angular power spectral density as a function of the emission angle $\theta$ and the particle-graphene separation $y_0$ for (a) TM mode or (b) TE mode. In (a, b), the insets plot the angular power spectral density and emission angle of the swift charged particle at $y_0 = 5$ nm and $y_0 = 50$ nm. (c) The radiation peak-intensity versus the particle-graphene separation $y_0$ for TE and TM modes.